\documentclass[aps,prb,twocolumn,nofootinbib,superscriptaddress]{revtex4-2}
\usepackage[]{graphicx,hyperref}
\usepackage{dcolumn}
\usepackage{bm}
\usepackage{amsmath}
\usepackage{braket}
\usepackage{physics}
\usepackage{amsfonts}
\usepackage{amssymb}
\usepackage{color}

\usepackage{multirow}
\newcommand{\nn}{\nonumber}

\hypersetup{
setpagesize=false,
 bookmarksnumbered=true,
 bookmarksopen=true,
 colorlinks=false,
 linkcolor=black,
 citecolor=blue,
}

\begin{document}
\preprint{APS/123-QED}
\title{Possible Realization of Topological Crystalline Superconductivity \\ {with Time-Reversal Symmetry} in UTe$_2$}
\author{Jushin Tei}
\email{tei@blade.mp.es.osaka-u.ac.jp}
\affiliation{Department of Materials Engineering Science, Osaka University, Toyonaka 560-8531, Japan}
\author{Takeshi Mizushima}
\affiliation{Department of Materials Engineering Science, Osaka University, Toyonaka 560-8531, Japan}
\author{Satoshi Fujimoto}
\affiliation{Department of Materials Engineering Science, Osaka University, Toyonaka 560-8531, Japan}
\date{\today}

\begin{abstract}
The recent measurement of the de Haas--van Alphen effect in the spin-triplet superconductor UTe$_2$ [D. Aoki {\it et al}., J. Phys. Soc. Jpn. {\bf 91}, 083704 (2022)] supports 
cylindrical electron and hole Fermi surfaces, which implies 
that UTe$_2$ is trivial as a 3D time-reversal-invariant topological superconductor.
Inspired by this observation, we investigate the possible realization of a topological crystalline superconductor protected by the crystalline symmetry of UTe$_2$.
We examine Majorana surface states protected by mirror and two fold rotational symmetries for all symmetry-allowed odd-parity pairing states with time-reversal symmetry and clarify the corresponding topological invariants.
It is found that topological crystalline superconductivity can be realized for all irreducible representations of odd-parity pairing states of UTe$_2$
even for cylindrical Fermi surfaces.
\end{abstract}

\maketitle

\section{Introduction}
The discovery of superconductivity in the heavy fermion system UTe$_2$ 
has attracted much attention because of the prospect of spin-triplet  topological superconductivity
~\cite{ran2019v365}.
The highlights are huge upper critical fields along all directions beyond the Pauli limit
~\cite{aoki2019v88,knebel2019v88,ran2019v15},
a field-reentrant behavior for the $H\parallel b$-axis
~\cite{knebel2019v88,ran2019v15}, 
and a tiny decrease of NMR Knight-shift below $T_c$
~\cite{ran2019v365,nakamine2019v88,doi:10.7566/JPSJ.91.043705}.
These behaviors, which strongly support
spin-triplet pairing states, are similar to U-based ferromagnetic superconductors, UGe$_2$~\cite{UGe2}, URhGe~\cite{URhGe}, and UCoGe~\cite{PhysRevLett.99.067006}, which
are deemed nonunitary Weyl superconductors with broken time-reversal symmetry~\cite{aoki2012v81,doi:10.7566/JPSJ.88.022001}.
An important difference of UTe$_2$ from these ferromagnetic superconductors is that it is paramagnetic
down to 25 mK~\cite{sundar2019v100}.
Although the possibility of nonunitary chiral pairing states with broken time-reversal symmetry has been discussed from 
STM measurements~\cite{STM}, Kerr-effect measurements~\cite{Kerr}, and theoretical studies~\cite{Shaffer2022v106,nevidomskyy2020stability}, solid evidence of time-reversal symmetry breaking is lacking to this date.
Thus it is legitimate to explore the possibility of a time-reversal-invariant topological superconducting state in UTe$_2$.
However, the symmetry of the gap function of UTe$_2$, 
which determines topological properties, remains controversial even within time-reversal-invariant unitary states.
Furthermore, the structure of the Fermi surface is crucially important for topological characters.
For superconductors with time-reversal symmetry, which is classified as class DIII in the AZ classification, 
the Fermi surface must be a closed two-dimensional surface for a nonzero topological invariant, i.e., three-dimensional (3D) winding number~\cite{PhysRevB.78.195125}.
{Although first-principles band calculations are useful for the understanding of strongly correlated electron systems,
there is no complete consensus on the topology of the Fermi surface in UTe$_2$}
\cite{doi:10.7566/JPSJ.88.103701,ishizuka2019v123,PhysRevLett.123.217002,doi:10.7566/JPSCP.29.011006,PhysRevB.100.134502}.

Recently, the observation of the de Haas--van Alphen (dHvA) effect in UTe$_2$ was successfully achieved by Aoki and co-workers~\cite{aoki2022v91}.
According to their result, the dHvA frequencies are compatible with cylindrical electron and hole Fermi surfaces.
This implies that the superconducting state of UTe$_2$ is, unfortunately, topologically trivial in the sense of 3D class DIII.
However, even for such open Fermi surfaces, the topological superconducting states protected by crystalline symmetry are still possible~\cite{PhysRevLett.106.106802,SnTe,morimoto2013v88,shiozaki2014v90,slager,doi:10.7566/JPSJ.82.113707,ueno2013v111,xiong2017v7}.
Motivated by this observation, in this paper we investigate the possible realization of topological crystalline superconductivity in UTe$_2$.
We classify topological invariants protected by the space group symmetry $Immm$ of UTe$_2$ and
examine surface Majorana zero-energy modes (MZMs) corresponding to the topological invariants.
It is found that the mirror symmetry and the two fold rotational symmetry play crucial roles in
the realization of topological crystalline superconducting states in  UTe$_2$.

The organization of this paper is as follows.
In Sec. II we briefly review the symmetry properties of all possible pairing states allowed by the point-group symmetry of  UTe$_2$, which is the basis of the following argument.
In Sec. III  we introduce topological invariants associated with crystalline symmetry, particularly focusing on one-dimensional (1D) crystalline winding numbers and the mirror Chern number. 
In Sec. IV we present an effective microscopic model for the band structure of UTe$_2$.
In Sec. V, using this model we investigate surface Majorana states of all irreducible representations of odd-parity pairing states and clarify topological invariants
arising from crystalline symmetry, which protect the surface Majorana zero-energy states.
The summary is given in Sec. VI.

\section{symmetry of superconductivity}
\begin{table}[b]
    \centering
    \caption{List of irreducible representations of $D_{2h}$ and basis functions of the $\bm{d}$ vector. $\hat{a}$, $\hat{b}$, and $\hat{c}$ are unit vectors parallel to the principal axes
    of the crystal structure. 
    $\mathcal{C}_{\mu}$ ($\mu=a,~b,~c$) is a two fold rotation with the rotation axis parallel to the $\mu$ axis.
    $\mathcal{M}_{\mu\nu}$ ($\mu,~\nu=a, ~b, ~c$) is a mirror reflection with the mirror plane parallel to the $\mu$ and $\nu$ axes.
    $\Delta^e$ is a real even function of a momentum $\bm{k}$.}
    \label{table:IR}
    \begin{tabular}{lcccccc|cc}
    \hline
    \multirow{2}{*}{IR} & \multicolumn{1}{l}{\multirow{2}{*}{$\mathcal{C}_a$}} & \multicolumn{1}{l}{\multirow{2}{*}{$\mathcal{C}_b$}} & \multicolumn{1}{l}{\multirow{2}{*}{$\mathcal{C}_c$}} & \multicolumn{1}{l}{\multirow{2}{*}{$\mathcal{M}_{bc}$}} & \multicolumn{1}{l}{\multirow{2}{*}{$\mathcal{M}_{ca}$}} & \multicolumn{1}{l|}{\multirow{2}{*}{$\mathcal{M}_{ab}$}} & \multicolumn{2}{c}{Basis function}                               \\ \cline{8-9} 
                        & \multicolumn{1}{l}{}                                                   & \multicolumn{1}{l}{}                                                   & \multicolumn{1}{l}{}                                                   & \multicolumn{1}{l}{}                                                        & \multicolumn{1}{l}{}                                                        & \multicolumn{1}{l|}{}                                                        & Orbital-triplet                            & Orbital-singlet     \\ \hline
    $A_u$               & $+$                                                                    & $+$                                                                    & $+$                                                                    & $-$                                                                         & $-$                                                                         & $-$                                                                         & $k_a\hat{a},~k_b\hat{b},~k_c\hat{c}$       & $i\Delta^e \hat{c}$ \\
    $B_{1u}$            & $-$                                                                    & $-$                                                                    & $+$                                                                    & $+$                                                                         & $+$                                                                         & $-$                                                                         & $k_b\hat{a},~k_a\hat{b}$ &                     \\
    $B_{2u}$            & $-$                                                                    & $+$                                                                    & $-$                                                                    & $+$                                                                         & $-$                                                                         & $+$                                                                         & $k_c\hat{a},~k_a\hat{c}$ & $i\Delta^e \hat{a}$ \\
    $B_{3u}$            & $+$                                                                    & $-$                                                                    & $-$                                                                    & $-$                                                                         & $+$                                                                         & $+$                                                                         & $k_c\hat{b},~k_b\hat{c}$ & $i\Delta^e \hat{b}$ \\ \hline
    \end{tabular}
\end{table}

We here present the symmetry classification of the possible pairing states of UTe$_2$, which is the basis of the following argument.
We assume that $f$ electrons of uranium sites are condensed into a spin-triplet odd-parity pairing state with time-reversal symmetry.
Then, the superconducting gap function is given by
\begin{eqnarray}
\Delta(\bm{k})=\bm{d}(\bm{k})\cdot\bm{\sigma}i\sigma_y,
\label{eq:BCSgap}
\end{eqnarray}
which satisfies
\begin{eqnarray}
i\sigma_y K\Delta(\bm{k})K(-i\sigma_y)=\Delta(-\bm{k}),
\label{eq:TRS}
\end{eqnarray}
where $\bm{d}$ is a $\bm{d}$ vector for spin-triplet pairing states, $\bm{\sigma}=(\sigma_x,\sigma_y,\sigma_z)$ is the Pauli matrices for spin degrees of freedom, and $K$ is the complex conjugate operator.
{More precisely, $\bm{\sigma}$ is defined not for spin-$1/2$, but for the Kramers doublet.
In general, atomic spin-orbit coupling is important for U-based compounds, and Cooper pairs are formed by electrons with the total angular momentum $j=5/2$.
However, we assume here pairings of electrons in the Kramers doublet states and hence we do not need to consider the $j=5/2$ states explicitly.}
The $\bm{d}$-vector is expressed in terms of the basis function of the point-group symmetry.     
UTe$_2$ has the body-centered orthorhombic lattice structure with the space-group symmetry
$Immm(\#71,D_{2h}^{25})$, and hence the relevant point group is $D_{2h}$, 
which has eight 1D irreducible representations (IRs).
Possible IRs for the $\bm{d}$ vector are $A_u$, $B_{1u}$, $B_{2u}$, and $B_{3u}$, as shown in Table~\ref{table:IR}.
Here we consider only the case that the direction of the $\bm{d}$-vector is constrained by the orbital degrees of freedom because of
strong spin-orbit interactions of $f$ electrons. 
Moreover, there are two types of uranium atoms in a unit cell. 
Thus two types of pairings of $f$ electrons are possible:
intraorbital pairings and interorbital pairings.
Each component of the $\bm{d}$ vector is represented by a matrix in this orbital space:
\begin{eqnarray}
\bm{d}(\bm{k})=
\begin{pmatrix}
\bm{d}_{11} & \bm{d}_{12} \\
 \bm{d}_{21} & \bm{d}_{22}
\end{pmatrix},
\label{eq:dmat}
\end{eqnarray}
where the indices $1$, $2$ denote the two $f$ orbitals.
{For odd-parity pairings, $\bm{d}$ vectors satisfy
\begin{eqnarray}
    \bm{d}_{11}(\bm{k}) = - \bm{d}_{22}(-\bm{k}) \\
    \bm{d}_{12}(\bm{k}) = - \bm{d}_{21}(-\bm{k})
\end{eqnarray}
For intraorbital pairings $\bm{d}_{11} = \bm{d}_{22}$, since the $\bm{d}$ vectors are odd functions of $\bm{k}$ due to Fermi statistics.
Also, for inter-orbital pairings $\bm{d}_{12}$ has both $\bm{k}$-odd and $\bm{k}$-even terms.
The former (latter) corresponds to orbital-triplet (orbital-singlet) pairings.}
It is noted that for preserving time-reversal symmetry (\ref{eq:TRS}), the gap function of the orbital-singlet state must be purely imaginary.
That is, the relative phase between the orbital-triplet pairs and the orbital-singlet pairs is $\pi/2$.
In Table ~\ref{table:IR} we summarize all possible basis functions for the $\bm{d}$-vector (\ref{eq:dmat}) 
allowed by the point-group symmetry.

\section{Topological invariants protected by crystalline symmetry}
\begin{figure}[tb]
    \centering
    \includegraphics[width=\linewidth]{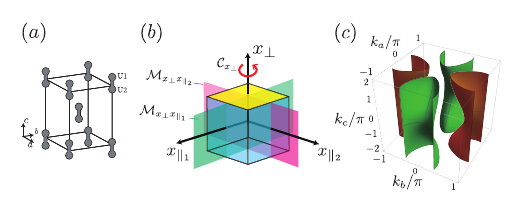}
    \caption{{(a) Schematic picture of the crystal structure of UTe$_2$. There are two types of uranium atoms in a unit cell.
    (b) Crystal axes, mirror symmetry planes, and a two fold rotational symmetry axis
    on an open surface perpendicular to $x_\perp$. The crystal axes $x_{\parallel_1}$ and $x_{\parallel_2}$
    are perpendicular to the $x_\perp$ axis. $x_\perp$ is one of either the $a$, $b$, and $c$ axis.
    (c) Cylindrical electron (orange) and hole (green) Fermi surfaces of UTe$_2$.
    }}
    \label{fig:ute2}
\end{figure}

In this section, we explain topological invariants related to surface MZMs protected by crystalline symmetry.
In Fig.~\ref{fig:ute2}(b) we show the geometrical configuration of the system used in the following argument.
The yellow surface in Fig.~\ref{fig:ute2}(b) is an open boundary surface of the crystal. 
The axis $x_\perp$ is normal to the surface, and $x_{\parallel_1}$ and $x_{\parallel_2}$ are orthogonal axes
parallel to the surface. 
$x_{\perp}$, $x_{\parallel_1}$, and $x_{\parallel_2}$ are, respectively, chosen to be one of the crystallographic axes $a$, $b$, or $c$ axes.
The surface possesses three crystalline symmetries:
two fold rotation $\mathcal{C}_{x_{\perp}}$ around the $x_{\perp}$-axis,
mirror reflection $\mathcal{M}_{x_{\perp}x_{\parallel_1}}$ with respect to the $x_\perp x_{\parallel_1}$ plane
and mirror reflection $\mathcal{M}_{x_{\perp}x_{\parallel_2}}$ with respect to the $x_\perp x_{\parallel_2}$ plane.
In the following, we see that such crystalline symmetries can be used 
for defining two topological invariants: the 1D crystalline winding number protected by chiral symmetry 
and the mirror Chern number.

We start with a general Bogoliubov--de Gennes (BdG) Hamiltonian $H(\bm{k})$ for superconducting states
\begin{eqnarray}
    \mathcal{H} = \frac{1}{2}\sum_{\bm{k},\alpha,\alpha'}
        \begin{pmatrix}
            c_{\bm{k}\alpha}^\dagger & c_{-\bm{k}\alpha} 
        \end{pmatrix}
        H(\bm{k})
        \begin{pmatrix}
            c_{\bm{k}\alpha'} \\
            c_{-\bm{k}\alpha'}^\dagger
        \end{pmatrix}
\end{eqnarray}
with
\begin{eqnarray}
    \label{BdG Hamiltonian}
        H(\bm{k}) = \begin{pmatrix}
            H_n(\bm{k}) & \Delta(\bm{k}) \\
            \Delta^\dagger(\bm{k}) & -H_n^T(-\bm{k})
        \end{pmatrix}
\end{eqnarray}
where $c_{\bm{k}\alpha}^\dagger (c_{\bm{k}\alpha})$ is the creation (annihilation) operator
of the electron with momentum $\bm{k}$,
$\alpha$ labels spin and orbital degrees of freedom,
$H_n(\bm{k})$ is the Hamiltonian for the normal state,
and $\Delta(\bm{k})$ is the gap function.
The BdG Hamiltonian has time-reversal symmetry (TRS) and particle-hole symmetry (PHS),
\begin{eqnarray}
    \Theta H(\bm{k})\Theta^{-1} &=& H(-\bm{k}), ~ \Theta = \begin{pmatrix}
    i\sigma_y  & \\
    & i\sigma_y
    \end{pmatrix} K, \\
    C H(\bm{k})C^{-1} &=& -H(-\bm{k}), ~ C = \begin{pmatrix}
        & 1\\
        1 & 
    \end{pmatrix}K,
\end{eqnarray}
Then we can define a chiral symmetry operator which anticommutes with the BdG Hamiltonian:
\begin{eqnarray}
    \Gamma = i\Theta C.
    \label{eq:gam}
\end{eqnarray}
With this chiral symmetry we can define the topological invariant referred to as the 3D winding number  for
3D class DIII superconductors with TRS.
However, as mentioned in the Introduction,  the 3D winding number vanishes for cylindrical Fermi surfaces.
Thus, instead, we consider topological invariants protected by crystalline symmetry.

We denote a crystalline symmetry operator as $U$.
Then the Hamiltonian for the normal state is transformed as
\begin{eqnarray}
    UH_n(\bm{k})U^\dagger = H_n(\hat{g}\bm{k}),
\end{eqnarray}
where $\hat{g}$ is the 3D representation of $U$ acting in the $\bm{k}$-space.
The gap function is transformed following the characters of each IR shown in Table~\ref{table:IR},
\begin{eqnarray}
    \label{eq:parity gap}
    U \Delta(\bm{k}) U^T = s \Delta(\hat{g}\bm{k}),
    \label{eq:ug}
\end{eqnarray}
where $s=+1$ or $-1$.
Then we can define the crystal symmetry operator acting on Nanbu space, 
\begin{eqnarray}
    \tilde{U}H(\bm{k})\tilde{U}^\dagger = H(\hat{g}\bm{k}),
\end{eqnarray}
with
\begin{eqnarray}
    \tilde{U} = \begin{pmatrix}
        U & \\
        & s U^*
    \end{pmatrix}.
\end{eqnarray}
For the point group symmetry $D_{2h}$, $U$ is $\mathcal{C}_{\mu}$ or $\mathcal{M}_{\mu\nu}$ with $\mu, \nu=a, b, c$, as shown in Table I.
With these settings, in the following we consider two topological invariants for topological crystalline superconductivity: 
the 1D crystalline winding number and the mirror Chern number.

\subsection{1D crystalline winding number}

Here we consider the 1D winding number for a 1D subspace of the 3D Brillouin zone (BZ).
The 1D winding number is defined for systems with chiral symmetry.
However, for the chiral symmetry operator for the class DIII, Eq.~(\ref{eq:gam}),  the 1D winding number always vanishes,
because it anticommutes with the time-reversal symmetry operator $\Theta$,  $\{ \Theta,~\Gamma\} = 0$.
Nevertheless, in the case with a crystalline symmetry, 
we can introduce another chiral symmetry operator $\Gamma_U$ defined as
\begin{eqnarray}
    \Gamma_U = e^{i\phi_U}\tilde{U}\Gamma,
\end{eqnarray}
where $e^{i\phi_U}$ is a phase factor which ensures $\Gamma_U^2 = 1$.
Then we define the 1D crystalline winding number as
\begin{eqnarray}
    \label{eq:sym winding num}
    w_U(\bm{k_{\parallel}}) = -\frac{1}{4\pi i}\int d\bm{k_{\perp}}
    \tr [\Gamma_U H^{-1}\partial_{\bm{k}_{\perp}}H(k_\perp)],
\end{eqnarray}
where $\bm{k}_{\perp}$ is the momentum in the 1D subspace, and $\bm{k}_{\parallel}$ is perpendicular to $\bm{k}_{\perp}$.
Both $\bm{k}_{\perp}$ and  $\bm{k}_{\parallel}$ are on a symmetry axis or a symmetry plane, satisfying $\bm{k} = \hat{g}\bm{k}$.
If the gap function does not change its sign under a crystalline symmetry operation $U$,
i.e., $s=+1$ in Eq. (\ref{eq:ug}), $\Gamma_U$ commutes with $\Theta$:
\begin{eqnarray}
    [\Theta, \Gamma_U]=0.
    \label{eq:gamUc}
\end{eqnarray}
By virtue of this property, similar to class BDI, $w_U(\bm{k_{\parallel}})$ does not vanish identically.
Therefore, for IRs with the character ``$+$" shown in Table~\ref{table:IR},
the 1D crystalline winding number defined  with the corresponding crystalline symmetry
can be nonzero and surface MZMs protected by crystalline symmetry can appear.
Note that the 1D winding number can be expressed in terms of quantities
on the Fermi surface~\cite{sato2011v83}. The Fermi surface formulas for $w_U(\bm{k}_{\parallel})$ are given in the Appendix. 
Then it is necessary for nonzero $w_U(\bm{k}_{\parallel})$ that the 1D subspace BZ crosses the Fermi surface.
In Sec.~V we will discuss the 1D crystalline winding number for each pairing state using the Fermi surface formulas.

\subsection{Mirror Chern number}

Another topological invariant can be defined on a mirror plane that is parallel to the $x_\perp$ and $x_{\parallel_\ell}$ ($\ell=1,~2$).
On the mirror plane the BdG Hamiltonian can be block diagonal in the diagonal basis 
of the mirror reflection operator $\tilde{\mathcal{M}}_{x_\perp x_{\parallel_{\ell}}}$.
Then, the mirror Chern number $\nu_{x_{\perp} x_{\parallel_{\ell}}}$ is defined as
\begin{eqnarray}
    \nu_{x_{\perp} x_{\parallel_{\ell}}}=\frac{1}{2}[\nu_{x_\perp x_{\parallel_{\ell}}}(+i)- \nu_{x_\perp x_{\parallel_{\ell}}}(-i) ],
\end{eqnarray}
\begin{eqnarray}
    \nu_{x_\perp x_{\parallel_{\ell}}}(\lambda) = \frac{1}{2\pi}\int_{\mathrm{BZ}} dk_\perp dk_{\parallel_{\ell}} ~
    \mathcal{F}_{x_\perp x_{\parallel_{\ell}}}^\lambda (\bm{k}),
\end{eqnarray}
where $\lambda=\pm i$ are the eigenvalues of $\tilde{\mathcal{M}}_{x_\perp x_{\parallel_{\ell}}}$,
and $\mathcal{F}_{x_\perp x_{\parallel_{\ell}}}^\lambda $ is the Berry curvature of the eigen states on the mirror plane in the BZ
with the mirror eigenvalue $\lambda$. 
Since we consider the cylindrical Fermi surfaces shown in Fig.~\ref{fig:ute2}(c),
the mirror Chern number can be nonzero only on the $\mathcal{M}_{ab}$ plane.
Furthermore, for the basis function of the $B_{2u}$ and $B_{3u}$ states shown in Table I, the mirror Chern number $\nu_{ab}$ vanishes.
Therefore surface MZMs protected by the mirror Chern number can appear only in the $A_u$ and $B_{1u}$ states.
We demonstrate it in Sec.~V.

\section{Effective model}

We here present an effective model of UTe$_2$ used for the following analysis.
For simplicity, we consider only the electron Fermi surface [Fig.~\ref{fig:ute2}(c), orange face].
The contribution of the hole Fermi surface can be similarly taken into account. 
Our simplified model is composed of two $f$-electron orbitals.

We neglect the staggered Rashba-type spin-orbit interaction due to local inversion symmetry breaking, which was discussed in previous studies~\cite{shishidou2021v103}.
{In general, by including the staggered Rashba-type spin-orbit interaction, 
there may be some mixing with even parity pairing.
However, the magnitude of the mixing is suppressed 
by the factor $E_{Rashba}/E_F$, where $E_{Rashba}$ is the energy scale of Rashba interaction,
unless the pairing interaction in the spin-singlet channel is comparable to that in the spin-triplet channel~\cite{fujimoto2007v76}
and is not expected to qualitatively affect the results.}
In fact, the energy scale of the staggered Rashba-type interaction is much smaller than other energy scales relevant to the band structure.

Then, following ref.~\cite{shishidou2021v103}, we assume the Hamiltonian for $f$ electrons in the normal state as
\begin{eqnarray}
\label{eq:Normal Hamiltonian}
        H_n(\bm{k}) = \epsilon_0(\bm{k}) - \mu+ f_x(\bm{k})\tau_x + f_y(\bm{k})\tau_y,  
\end{eqnarray}
with
\begin{eqnarray}
    \epsilon_0(\bm{k}) &=& 2t_1 \cos k_a + 2t_2 \cos k_b, \\
    f_{x}(\bm{k}) &=& t_3 + t_4 \cos(k_a/2)\cos(k_b/2)\cos(k_c/2), \\
    f_{y}(\bm{k}) &=& t_5 \cos(k_a/2)\cos(k_b/2)\sin(k_c/2),
\end{eqnarray}
where $\tau_{x, y}$ are Pauli matrices for orbital degrees of freedom.
To reproduce a cylindrical electron Fermi surface,
we choose the parameters as follows:
$\mu = -1.8,~ t_1 = -0.5,~ t_2 = 0.375,~ t_3 = -0.7, ~ t_4 = 0.65, ~ t_5 = -0.65$.
The model has two energy bands $E_{\pm} = \varepsilon - \mu \pm|f|$, with 
$f = f_x - i f_y$. 
The Fermi level crosses the lower band $E_{-}$ for which a cylindrical electron Fermi surface is realized as depicted in Fig.~\ref{fig:ute2}(c).

In the superconducting state, the gap functions are given by Eqs.~(\ref{eq:BCSgap}) and (\ref{eq:dmat}) as discussed in Sec.~II.
In numerical calculations for surface states presented in the next sections, to simplify the analysis we consider only interorbital pairing states
which consist of orbital-triplet and orbital-singlet pairings,
\begin{eqnarray}
    \Delta_{12}(\bm{k}) = i \bm{d}_{12}\cdot\boldsymbol{\sigma}\sigma_y.
\end{eqnarray}
In fact, we have examined that the inclusion of intraorbital pairing states does not change the qualitative features of the results (see Appendix).

We now comment on the gap-node structure.
{In the case of cylindrical Fermi surfaces shown in Fig.~\ref{fig:ute2}(c),
the $A_u$ and $B_{1u}$ states are fully gapped
while the $B_{2u}$ and $B_{3u}$ states are
Dirac superconducting states with point nodes
on the lines $k_a = 0,~k_c=0$ and $k_b = 0,~k_c=0$ respectively.
Note that these point nodes are protected by two fold rotational symmetry.}
In this paper we call both fully gapped superconductors and nodal superconductors with surface  MZMs protected by crystalline symmetry 
topological crystalline superconductors.

\section{Majorana zero-energy surface states and the corresponding topological invariants}

In this section we present numerical results of surface MZMs and topological invariants which protect them.
For numerical calculations of quasiparticle energy spectra, we use a system with open boundary surfaces perpendicular to the $x_{\perp}$ axis
($x_{\perp}=a, b, c$).
The system size $L$ along $x_{\perp}$ is set as $L = 50$ for all calculations.
As shown below, surface MZMs protected by crystalline topological invariants
appear for all IRs of odd-parity pairing states even for the cylindrical Fermi surfaces.
In Secs. V A, V B, V C, and V D, we mainly discuss the case only with the electron Fermi surface.
The results in the case with the hole Fermi surface can be deduced from the results of the electron surface.
The summary of surface MZMs in the case with both the electron and hole Fermi surfaces is given in Table~\ref{table:result}.

\subsection{$A_u$ pairing state}
\begin{figure*}[t]
    \includegraphics[width=\linewidth]{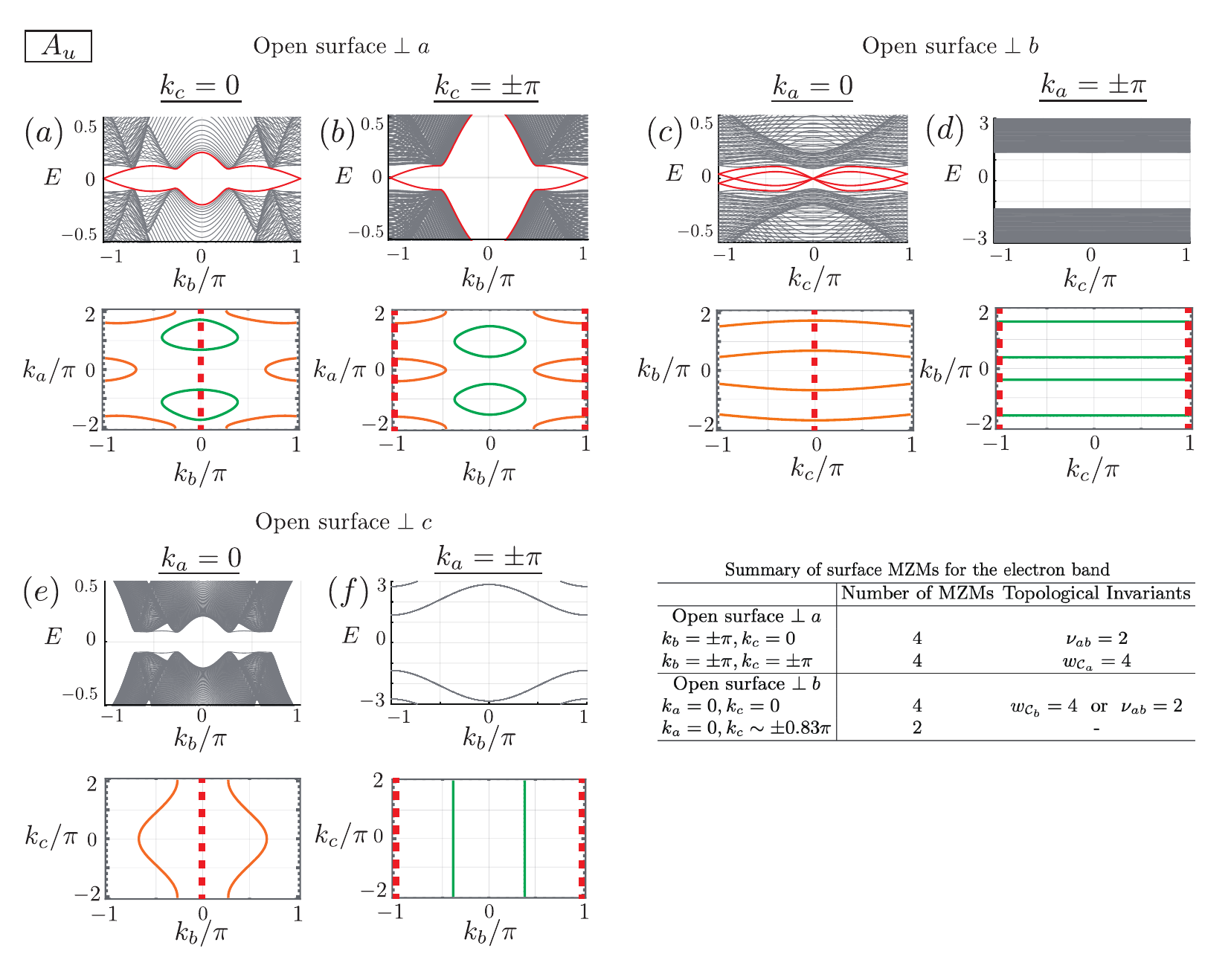}
    \caption{(a)--(f) Top:  Quasiparticle energy spectra of the $A_u$ state for open surfaces perpendicular to the $x_{\perp}$ axis ($x_{\perp}=a, b, c$). 
    The system size along the $x_{\perp}$ axis used for the calculations is $L=50$.
    The calculations are carried out for the model with the cylindrical electron Fermi surface, and
    the contributions from the hole band are not included.
    Red curves correspond to low-energy surface states.
    The gapless surface states appear in (a), (b), and (c). 
    Bottom: The cross sections of the electron Fermi surface (orange) and the hole Fermi surface (green) at (a) $k_c=0$, (b) $k_c=\pm\pi$, (c) $k_a=0$, (d) $k_a=\pm\pi$, (e) $k_a=0$, and (f) $k_a=\pm\pi$.
    The 1D crystalline winding number can be nonzero when the axis of $\mathcal{C}_{x_\perp}$ rotation crosses the Fermi surfaces.
    Red broken lines are $a$-, $b$- and $c$-rotational axes in (a)--(f).
    In the lower-right table, we show the summary of the degeneracy of the spectrum and the corresponding topological invariants for the electron band.
    In the middle column, ``Number of MZMs" means the degeneracy of MZMs protected by the 1D crystalline winding number, 
    or the total number of zero-energy states associated with the mirror Chern number.
    }
    \label{fig:Au}
\end{figure*}

We start with the $A_u$ pairing state. We use the following form of the $d$-vector.
\begin{eqnarray}
    \bm{d}_{A_u}\bm{(k)} = \begin{pmatrix}
        C_1 \sin k_a \\
        C_2 \sin k_b \\
        C_3 \sin k_c + i\Delta^e
    \end{pmatrix} 
\end{eqnarray}
For numerical calculations of surface states, we set $C_1 = C_2 = C_3 = \Delta^e = 0.1$ and assume that $\Delta^e$ is a constant for simplicity.
We show quasiparticle energy spectra for the $A_u$ state obtained 
by diagonalizing the BdG Hamiltonian with open boundary surfaces in Fig.~\ref{fig:Au}.
We note that in these numerical calculations, only the electron Fermi surface is taken into account.
Surface MZMs in the case with the hole Fermi surface can be deduced from the results of the electron Fermi surface.

First we discuss the results for the open surface perpendicular to the $a$-axis ($x_{\perp}=a$) shown in Figs.~\ref{fig:Au}(a) (top) and \ref{fig:Au}(b) (top).
In Fig.~\ref{fig:Au}(b) (top), we see four fold-degenerate MZMs at $k_b = \pm \pi$, $k_c = \pm \pi$.
In fact, these surface MZMs are protected by crystalline symmetry.
The points $k_b = \pm \pi$, $k_c = \pm \pi$ are invariant under the $a$ rotation $\mathcal{C}_a$.
{Note that the points $k_b = \pm \pi$, $k_c = \pm \pi$ are changed to $k_b = \mp \pi$, $k_c = \mp \pi$ by $\mathcal{C}_{a}$, 
and the transformed points are connected to the original points by reciprocal lattice vectors.}
The $A_u$ state has the character ``$+$" for this rotation (see Table I).
Thus, as discussed in Sec.~III. A, the 1D crystalline winding number can be defined at these $k$-points.
For a 1D subspace along the $a$ axis, the 1D crystalline winding number (\ref{eq:sym winding num}) can be reduced to
the Fermi surface formula (see Appendix A), 
\begin{eqnarray}
    w_{\mathcal{C}_{a}} = \sum_{E(\bm{k}_F)=0}\mathrm{sgn}[(\bm{d}_{12})_ a]\mathrm{sgn}[\partial_{k_a}E_-],
    \label{eq:wd_ca}
\end{eqnarray}
which is calculated from information on the Fermi surface.
As seen by the red broken line in Fig.~\ref{fig:Au}(b) (bottom), 
the 1D lines with $k_b=\pm\pi$, $k_c=\pm\pi$ in the momentum space
cross the Fermi surfaces four times. 
Then we find 
\begin{eqnarray}
    w_{\mathcal{C}_a}(k_b=\pm\pi,~k_c=\pm\pi)=4.
\end{eqnarray}
As a result, four fold-degenerate surface MZMs appear, in agreement with the numerical result.
It is noted that although $k_b = k_c = 0$ is also invariant under the $a$ rotation, MZMs do not appear since there is no Fermi surface crossed by the 1D line
at $k_b = k_c = 0$, and the 1D crystalline winding number is zero.
However, if one takes into account the hole Fermi surface [Fig.~\ref{fig:Au}(a) (bottom) green line], which has cross sections with
the   $k_b = k_c =0$  red broken line, 
$w_{\mathcal{C}_a}=4$ is obtained and then surface MZMs appear.

On the other hand, in Fig.~\ref{fig:Au}(a) (top), the surface MZMs appear at $k_b = \pm \pi$, $k_c=0$.
Since the 1D lines with $k_b = \pm\pi$, $k_c=0$ are not invariant under the $a$ rotation,
one needs to consider other crystalline symmetry protecting them.
In fact,  these 1D lines are on the mirror plane of $\mathcal{M}_{ab}$, i.e., $k_c=0$,  and hence 
surface MZMs protected by the mirror Chern number are possible to occur.
Actually, we obtain the nonzero mirror Chern number for the $k_c=0$ plane as 
$\nu_{ab} = 2$. 
This implies the existence of two doubly-degenerate surface MZMs, which is in perfect agreement with the numerical result.
However, we should note that this result may be affected by the hole Fermi surface.
If the gap function of the hole band is the same as that of the electron band,
the mirror Chern number of the electron band cancels with that of the hole band.
This means that pair hopping processes between these two bands may generate an energy gap of Majorana surface states.
On the other hand, if the relative sign between the gap functions of the two bands is minus, the cancellation does not occur, and
the surface MZMs survive.

\begin{figure}[t]
    \includegraphics[width=\linewidth]{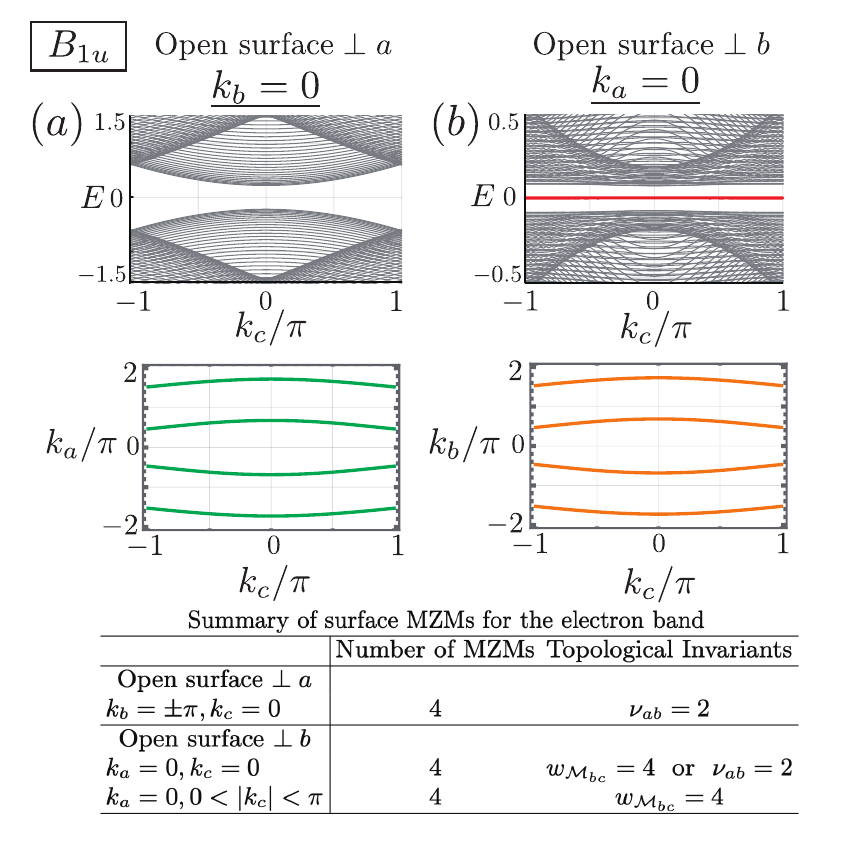}
    \caption{Top: Quasiparticle energy spectra of the $B_{1u}$ state for an open surface perpendicular to the $a$ axis (a) and the $b$ axis (b).
    The results for the electron band are shown.
    Bottom:  The cross section of the electron Fermi surface at $k_a=0$ (orange) and the hole Fermi surface at $k_b=0$ (green).
    The 1D crystalline winding number can be nonzero when the mirror plane crosses the Fermi surface.
    In the lower table we show the summary of the degeneracy of the spectra and the corresponding topological invariants for the electron band.
    In the middle column, ``Number of MZMs" means the degeneracy of MZMs protected by the 1D crystalline winding number, 
    or the total number of zero-energy states associated with the mirror Chern number.
    }
    \label{fig:B1u}
\end{figure}

In the case with an open surface perpendicular to the $b$ axis ($x_{\perp}=b$),
four fold-degenerate surface MZMs appear at $k_a = k_c = 0$, as shown in Fig.~\ref{fig:Au}(c) (top).
The 1D line at $k_a = k_c = 0$ is invariant under the $b$ rotation $\mathcal{C}_{b}$, for which the character of the $A_u$ state is ``+".
We obtain $w_{\mathcal{C}_b}=4$, which is in agreement with the numerical result of the surface MZMs. 
It is noted that on the $k_c=0$ plane, the mirror Chern number is also nonzero, $\nu_{ab} = 2$.
This means that the four MZMs at $k_c=0$ are protected also by the mirror symmetry.
In Fig.~\ref{fig:Au}(c) (top), we also see surface MZMs at  $k_a = 0$, $k_c \sim \pm 0.83\pi$, which are not on a symmetric axis.
The mirror Chern number on the $k_a=0$ plane is zero because the cross sections of the Fermi surface with  the $k_a=0$ plane
is a quasi-1D open Fermi surface.
Thus there is no topological invariant that protects these MZMs.
We speculate that the MZMs at $k_a = 0$, $k_c \sim \pm 0.83\pi$ are accidental ones, arising from a simplification of the model Hamiltonian,
and gapped out by adding symmetry-allowed terms to the Hamiltonian.
In Fig.~\ref{fig:Au}(d) (top), MZMs do not appear at $k_a = \pm\pi$, $k_c=\pm\pi$, which are also invariant under $b$ rotation $\mathcal{C}_b$, since there is no Fermi surface.
However, if one takes into account the hole Fermi surface [Fig.~\ref{fig:Au}(d) (bottom) green line], which has cross sections with
the $k_a = \pm \pi$, $k_c =\pm \pi$ red broken lines, 
$w_{\mathcal{C}_a}=4$ is obtained and then surface MZMs appear.

In the case with an open surface perpendicular to the $c$-axis ($x_{\perp}=c$),
there is no surface MZM, as shown in Figs.~\ref{fig:Au}(e) (top) and \ref{fig:Au}(f) (top).
In fact, the 1D crystalline winding number vanishes, because no Fermi surface crosses the $c$ rotation axis.
The existence of the hole Fermi surface does not change this situation.

The results in the case with both the electron and hole Fermi surfaces are summarized in Table II.

\subsection{$B_{1u}$ pairing state}
Next, we present the results of the $B_{1u}$ state.
The $\bm{d}$ vector is
\begin{eqnarray}
    d_{B_{1u}}(\bm{k}) = \begin{pmatrix}
        C_1 \sin k_b \\
        C_2 \sin k_a \\
        0
    \end{pmatrix}.
\end{eqnarray}
We set $C_1 = C_2 = 0.1$ for numerical calculations.
The $\bm{d}$ vector does not depend on $k_c$. Thus surface MZMs do not appear in the open surface perpendicular to the $c$ axis.
In Fig.~\ref{fig:B1u} we show the calculated results of the energy spectra for the electron Fermi surface in the case of
$x_{\perp}=a$ and $x_{\perp}=b$.
We find the surface Majorana zero-energy flat band at $k_a=0$ for the surface perpendicular to the $b$ axis.
The surface flat band is protected by mirror symmetry with the mirror plane $k_a=0$.
In fact, we can define the 1D crystalline winding number by putting $U=\mathcal{M}_{bc}$ in Eq.~(\ref{eq:sym winding num}).
Note that the character of $\mathcal{M}_{bc}$ for the $B_{1u}$ state is ``+," as shown in Table I.
Then, on the $k_a = 0$ plane, the 1D crystalline winding number $w_{\mathcal{M}_{bc}}(k_c)$ can be nonzero.
The Fermi surface formula of $w_{\mathcal{M}_{bc}}(k_c)$ is given by
\begin{eqnarray}
    w_{\mathcal{M}_{bc}}(k_c) = \sum_{E(k_b) =0}\mathrm{sgn}[d_a(k_b)]\mathrm{sgn}[\partial_{k_b}E_-(k_b)].
\end{eqnarray}
For any values of $k_c$, 1D momentum subspace along the $b$ axis on the $k_a = 0$ plane crosses the Fermi surface four times 
and hence $w_{\mathcal{M}_{bc}}(k_c)=4$, which is in agreement with the four fold degeneracy of the flat band. 
At $k_c=0$, which is the $\mathcal{M}_{ab}$ mirror plane, the mirror Chern number can be defined, and  actually, we obtain
$\nu_{ab}=2$. However, according to the numerical results of the surface energy spectrum,
there is no MZM at $k_c=0$ in addition to the MZM flat band shown in Fig.~\ref{fig:B1u}(b) (top).
Thus the MZMs at $k_c=0$ in the flat band are protected by both $w_{\mathcal{M}_{bc}}$ and $\nu_{ab}$.
On the other hand, there is no surface MZM on the mirror plane of $\mathcal{M}_{ca}$ for the surface perpendicular to the $a$ axis,
since there is no Fermi surface on the mirror plane.
If one takes into account the hole Fermi surface, 
a similar surface MZM flat band will appear on the $k_b = 0$ plane which is the mirror plane of $\mathcal{M}_{ca}$.
The results in the case with both the electron and hole Fermi surfaces are summarized in Table II.

\subsection{$B_{2u}$ pairing state}
\begin{figure}[t!]
    \includegraphics[width=\linewidth]{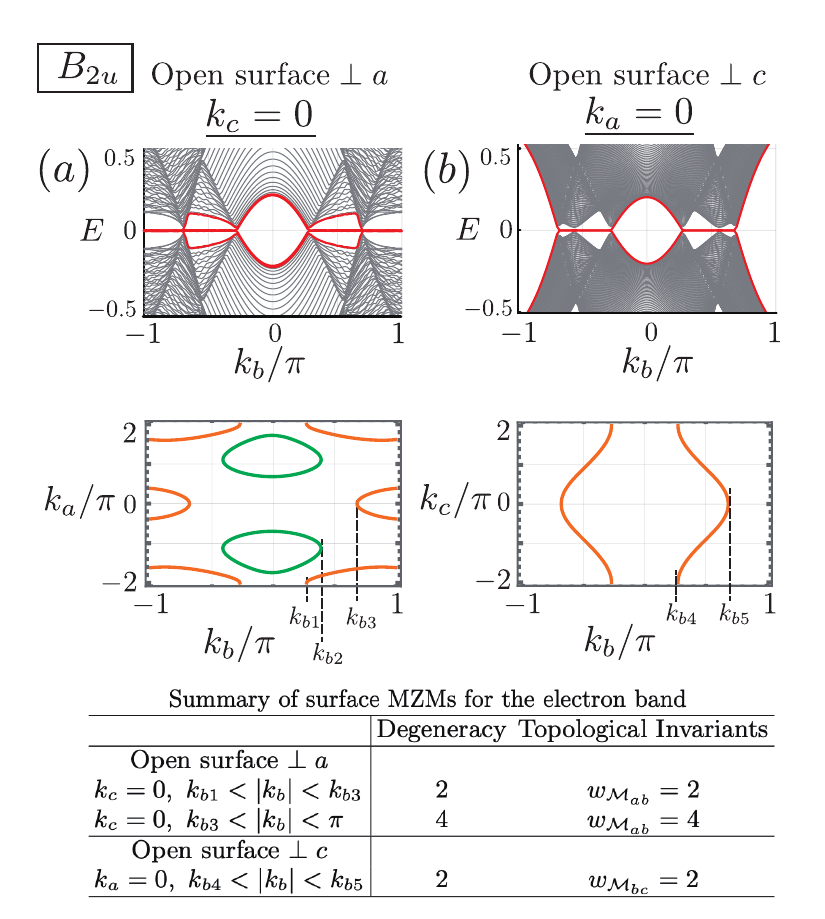}
    \caption{Top: Quasiparticle energy spectra of the $B_{2u}$ state for open surfaces perpendicular to 
    the $a$ axis (a) and the $c$ axis (b). 
    The results for the electron band are shown.
    Bottom: The cross section of the Fermi surfaces at $k_c=0$ (a) and $k_a=0$ (b). 
    The orange (green) curves are the electron (hole) Fermi surfaces. 
    In the lower table we show the summary of the degeneracy of the flat band and the corresponding topological invariants for the electron band.
   }
    \label{fig:B2u}
\end{figure}
The $\bm{d}$ vector of the $B_{2u}$ state is given by,
\begin{eqnarray}
    d_{B_{2u}}(\bm{k}) = \begin{pmatrix}
        C_1 \sin k_c + i\Delta^e \\
        0 \\
        C_2 \sin k_a 
    \end{pmatrix}.
\end{eqnarray}
We set $C_1 = C_2 = \Delta^e = 0.1$ for numerical calculations.
In this case, since the $\bm{d}$ vector does not depend on $k_b$, surface MZMs can appear for $x_{\perp}=a$ and $x_{\perp}=c$.
In Fig.~\ref{fig:B2u} numerical results of the energy spectra for the electron Fermi surface in the case with these open surfaces are shown.
In both cases surface Majorana flat bands appear.
As in the case of the $B_{1u}$ state, these flat bands are associated with the 1D crystalline winding number protected by mirror symmetry with respect to
the mirror planes $k_c=0$ and $k_a=0$.
As mentioned in previous sections, the 1D crystalline winding number is determined by the number of crossing points of the Fermi surface and
a 1D subspace where the topological invariant is defined.

\begin{figure}[t]
    \includegraphics[width=\linewidth]{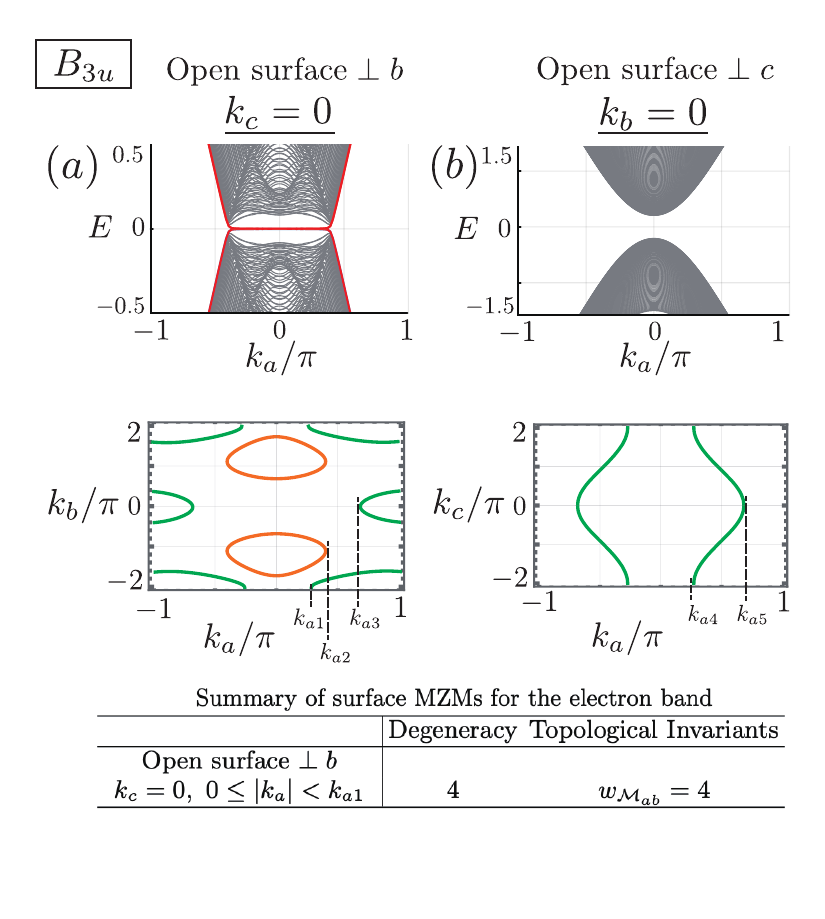}
    \caption{Top: Quasiparticle energy spectra for the $B_{3u}$ state with open surfaces perpendicular to 
    the $b$ axis (a) and the $c$ axis (b).
    The results for the electron band are shown.
    Bottom: The cross section of the Fermi surfaces at $k_c=0$ (a) and $k_b=0$ (b).
    The orange (green) curves are the electron (hole) Fermi surfaces.
    In the lower table we show the summary of the degeneracy of the flat band and the corresponding topological invariants for the electron band.
    }
    \label{fig:B3u}
\end{figure}

For the $k_c=0$ plane [see Fig.~\ref{fig:B2u}(a)], we find $w_{\mathcal{M}_{ab}}(k_b)=2$ for $k_{b1}<|k_b|<k_{b3}$
and $w_{\mathcal{M}_{ab}}(k_b)=4$ for $k_{b3}<|k_b|<\pi$, where the definitions of $k_{b1}$ and $k_{b3}$ are given in Fig.~\ref{fig:B2u}(a) (bottom). 
These results are in agreement with the degeneracy of the flat band obtained by the numerical calculation.
For the $k_a=0$ plane [see Fig.~\ref{fig:B2u}(b)], we find $w_{\mathcal{M}_{bc}}(k_b)=4$ for $k_{b4}<|k_b|<k_{b5}$, where the definitions of $k_{b4}$ and $k_{b5}$ are given in Fig.~\ref{fig:B2u}(b) (bottom). 
The results of the topological invariants are in agreement with the degeneracy obtained by the numerical calculation.
The inclusion of the hole Fermi surface affects the results for $x_{\perp}=a$ [see Fig.~\ref{fig:B2u}(a) (bottom)].
Both the electron and hole Fermi surfaces contribute to the 1D winding number for $k_{b1} \leq |k_b| \leq k_{b2}$, where the definition of $k_{b2}$ is given in Fig.~\ref{fig:B2u} (a)(bottom). 
The 1D momentum subspace for $k_{b1} \leq |k_b| \leq k_{b2}$ crosses  the electron Fermi surface twice and the hole Fermi surface four times.
Thus the flat band still appears, but the degeneracy is changed.
The degeneracy depends on the relative sign between the gap functions of the electron and hole bands.
The results in the case with both the electron and hole Fermi surfaces are summarized in Table II.

\subsection{$B_{3u}$ pairing state}

The $d$ vector of the $B_{3u}$ state  is given by
\begin{eqnarray}
    d_{B_{3u}}(\bm{k}) = \begin{pmatrix}
        0 \\
        C_1 \sin k_c + i\Delta^e \\
        C_2 \sin k_b 
    \end{pmatrix}.
\end{eqnarray}
We set $C_1 = C_2 = \Delta^e = 0.1$ for numerical calculations.
In this case, since the $\bm{d}$ vector does not depend on $k_a$, surface MZMs can appear for $x_{\perp}=b$ and $x_{\perp}=c$.
In Fig.~\ref{fig:B3u} numerical results of the energy spectra for the electron Fermi surface in the case with these open surfaces are shown.
Note that bulk nodes appear at $k_a \neq 0$ in Fig.~\ref{fig:B3u}(a), but these are for the simplicity of our model.
The stable point nodes are located on the hole Fermi surface.

We find surface MZMs, which constitute a flat band, for $x_{\perp}=b$.
The flat band is protected by the 1D crystalline winding number associated with mirror symmetry for $\mathcal{M}_{ab}$.
Using the Fermi surface formula of $w_{\mathcal{M}_{ab}}(k_a)$, we obtain $w_{\mathcal{M}_{ab}}(k_a)=4$ for $|k_a| < k_{a2}$, which is in agreement with the degeneracy
found in the numerical calculations of the energy spectrum.

The surface Majorana states in the case with the hole Fermi surface can be easily deduced from
the above results for the electron Fermi surface. 
The hole Fermi surface shown in Fig.~\ref{fig:ute2}(c) crosses the $\mathcal{M}_{ca}$ mirror plane, $k_b=0$.
Then we obtain the 1D crystalline winding number $w_{\mathcal{M}_{ca}}(k_a)=2$, and hence
a surface Majorana zero-energy flat band appears for the open surface with $x_{\perp}=c$.
The results in the case with both the electron and hole Fermi surfaces are summarized in Table II.
Note that, as seen from the shape of the Fermi surfaces which is cylindrical along the $c$ direction, 
the results for $B_{2u}$ and $B_{3u}$ are qualitatively similar.

\begin{table}[]
    \caption{Summary of surface MZMs of the model for UTe$_2$ with both the cylindrical electron and hole Fermi surfaces.
    The $\bm{d}$ vectors are assumed to be of the form $\sin k_{\mu}$.
    $k_{bj}$ and $k_{aj }$ with $j=1,2,3,4,5$ in the left columns of the $B_{2u}$ and $B_{3u}$ cases
    are defined in Figs.~\ref{fig:B2u} and \ref{fig:B3u}.
    In the right column, ``number of MZMs" means the degeneracy of MZMs protected by the 1D crystalline winding number, 
    or the total number of zero-energy states associated with the mirror Chern number.
    Most surface MZMs are protected by the 1D crystalline winding number $w_{U}$ defined on crystalline symmetry lines.
    In the cases of the $A_u$ and $B_{1u}$ states, the sum of the mirror Chern number of the electron and hole bands is $4$ or $0$, depending on
    the relative sign between the gap of the electron and hole surfaces. 
    Also, in the case of the $A_u$ and $B_{1u}$ state, some surface MZMs are protected by both the 1D crystalline winding number and the mirror Chern number,
    and the total number of the associated MZMs depends on the details of the model.
    }
    \label{table:result}
    \scalebox{0.85}{
    \begin{tabular}{lcc}
    \hline
                                    & Topological invariants~ & Number of MZMs\\ \hline \hline
    \multicolumn{1}{c}{$A_u$}          &                   \\ \hline
    \multicolumn{1}{c}{Open surface $\perp$ $a$} &                   \\
    $k_b = 0, k_c = 0$                      & $w_{\mathcal{C}_a} = 4$     & 4       \\
    $k_b = \pm\pi, k_c = \pm\pi$                  & $w_{\mathcal{C}_a} = 4$     & 4         \\ 
    $k_c=0$ plane                           & $\nu_{ab}=4$ or $0$ & $8$ or $0$ \\ \hline
    \multicolumn{1}{c}{Open surface $\perp$ $b$} &                   \\
    $k_a = 0, k_c = 0$                  & $w_{\mathcal{C}_b} = 4$       & 4     \\
    $k_a = \pm\pi, k_c = \pm\pi$              & $w_{\mathcal{C}_b} = 4$    & 4  \\
    $k_c=0$ plane                       & $\nu_{ab}=4$ or $0$  & $8$ or $0$ \\ \hline \hline
    \multicolumn{1}{c}{$B_{1u}$}         &                   \\ \hline
    \multicolumn{1}{c}{Open surface $\perp$ $a$} &  \\
    $k_b = 0, 0 \leq |k_c| < \pi$           & $w_{\mathcal{M}_{ca}} = 4$ & 4-fold flat band \\
    $k_c = 0$ plane                     & $\nu_{ab}=4$ or $0$ & $8$ or $0$ \\ \hline
    \multicolumn{1}{c}{Open surface $\perp$ $b$} \\
    $k_a = 0,0 \leq |k_c| < \pi$            & $w_{\mathcal{M}_{ca}} = 4$ & 4-fold flat band \\
    $k_c=0$ plane                       & $\nu_{ab}=4$ or $0$  & $8$ or $0$ \\ \hline \hline
    \multicolumn{1}{c}{$B_{2u}$}         &                   \\ \hline
    \multicolumn{1}{c}{Open surface $\perp$ $a$} &  \\
    $k_c = 0, 0 \leq |k_b| < k_{b1}$  & $w_{\mathcal{M}_{ab}} = 4$ & 4-fold flat band \\
    $k_c = 0, k_{b1} < |k_b| < k_{b2}$ & $w_{\mathcal{M}_{ab}} = 2$ or $6$ & 2 or 6-fold flat band \\
    $k_c = 0, k_{b2} < |k_b| < k_{b3}$ & $w_{\mathcal{M}_{ab}} = 2$ & 2-fold flat band \\
    $k_c = 0, k_{b3} < |k_b| < \pi$ & $w_{\mathcal{M}_{ab}} = 4$ & 4-fold flat band \\ \hline
    \multicolumn{1}{c}{Open surface $\perp$ $c$} &  \\
    $k_a = 0, k_{b4} < |k_b| < k_{b5}$ & $w_{\mathcal{M}_{bc}} = 2$ & 2-fold flat band \\ \hline \hline
    \multicolumn{1}{c}{$B_{3u}$}         &                   \\ \hline
    \multicolumn{1}{c}{Open surface $\perp$ $b$} &  \\
    $k_c = 0,0 \leq |k_a| < k_{a1}$ & $w_{\mathcal{M}_{ab}} = 4$ & 4-fold flat band \\
    $k_c = 0,k_{a1} < |k_a| < k_{a2}$ & $w_{\mathcal{M}_{ab}} = 2$ or $6$ & 2 or 6-fold flat band \\
    $k_c = 0,k_{a2} < |k_a| < k_{a3}$ & $w_{\mathcal{M}_{ab}} = 2$ & 2-fold flat band \\
    $k_c = 0,k_{a3} < |k_a| < k_{a4}$ & $w_{\mathcal{M}_{ab}} = 4$ & 4-fold flat band \\ \hline
    \multicolumn{1}{c}{Open surface $\perp$ $c$} &  \\
    $k_b = 0, k_{a4} < |k_a| < k_{a5}$ & $w_{\mathcal{M}_{ca}} = 2$ & 2-fold flat band \\ \hline 
    \end{tabular}
    }
\end{table}

\section{Summary}

Inspired by the recent experimental observation of the cylindrical Fermi surfaces of UTe$_2$, which implies that the superconducting state is trivial
in the sense of a 3D class DIII superconductor, 
we have examined the possibility of realizing topological crystalline superconductivity in UTe$_2$.
We have investigated numerically surface Majorana states for all IRs of odd-parity pairing states
and clarified topological invariants arising from crystalline symmetry which protect surface MZMs.
It is found that for all IRs, surface MZMs protected by topological invariants appear, supporting
the realization of topological crystalline superconductivity.
{The list of surface MZMs for all IRs is given in Table~\ref{table:result}.
For $A_u$ representation, the 1D crystalline winding number associated with two fold rotational symmetry, as well as the mirror Chern number play crucial roles in the stability of surface MZMs.
For $B_{1,2,3u}$ representation, the 1D crystalline winding number associated with mirror reflection symmetry gives rise to flat bands of surface MZMs.}

We note that these results are consistent with the analysis based on symmetry indicators~\cite{Kruthoff2017v7,SI1,doi:10.1126/sciadv.aaz8367,ono_shiozaki} and
the Atiyah-Hirzebruch spectral sequence (AHSS)~\cite{shiozaki_gomi_sato,shiozaki}.
The AHSS analysis predicts the existence of six (two) winding numbers in a 1D subspace and three (one) mirror Chern numbers for 
the $A_u$ ($B_{1,2,3u}$) representation of the space group $Immm$~\cite{shiozaki}. 
Our results demonstrate that these topological invariants are indeed the origins of surface MZMs obtained by numerical calculations.

{Finally, we briefly comment on the possible implications of our results for the case with 3D Fermi surface pockets.
Some recent studies discuss the possible existence of 3D Fermi pockets in addition to cylindrical Fermi surfaces in UTe$_2$~\cite{choi2023correlated,broyles2023revealing}.
In that case the 3D winding number of class DIII can be nonzero, and surface MZMs associated with the 3D winding number and those protected by crystalline symmetry discussed in this paper can coexist.
Even in such a case, the argument on crystalline topological invariants developed in this paper is useful for clarifying the origin of surface MZMs.}

\begin{acknowledgments}
The authors are grateful to K. Shiozaki for fruitful discussions, and providing his results of the AHSS analysis.
This work was supported by JST CREST Grant No.JPMJCR19T5, Japan, a Grant-in-Aid for Scientific Research on Innovative Areas, Quantum Liquid Crystals (JP22H04480) from JSPS of Japan, and JSPS KAKENHI (Grants No.~JP20K03860, No.~JP20H01857, No.~JP21H01039, and No.~JP22H01221).
\end{acknowledgments}

\begin{appendix}
\section{Fermi surface formulas for the 1D crystalline winding number}
We now derive useful formulas of the 1D crystalline winding number,
which are expressed only in terms of information on the Fermi surface, in the case of
our two-orbital model given in Sec. IV.
We follow the approach developed in Ref. ~\cite{sato2011v83} for single-orbital models.
We, particularly focus on the 1D crystalline winding number associated with the $a$ rotation in the case of the $A_u$ state.
The Fermi surface formulas for other cases can be derived in a similar manner.

The chiral symmetry operator associated with the $a$ rotation $\mathcal{C}_a$ is given by
\begin{eqnarray}
    \Gamma_{\mathcal{C}_{a}} &=& e^{i\phi_{\mathcal{C}_{a}}}\begin{pmatrix}
        & \mathcal{C}_{a} i\sigma_y \\
        \mathcal{C}_{a}^*i\sigma_y & 
    \end{pmatrix},
\end{eqnarray}
where $\mathcal{C}_a$ flips spin in the $b$ and $c$ direction
and exchanges two uranium sites in the unit cell.
$\mathcal{C}_a$ is explicitly written as
\begin{eqnarray}
    \mathcal{C}_{a} = i \sigma_x \tau_x .
\end{eqnarray}
We can diagonalize $\Gamma_{\mathcal{C}_{a}}$, 
\begin{eqnarray}
    V_\Gamma \Gamma_{\mathcal{C}_{a}} V_\Gamma^\dagger = \begin{pmatrix}
        \hat{1} & \\
        & -\hat{1}
    \end{pmatrix}, ~
    V_\Gamma = \frac{1}{\sqrt{2}}\begin{pmatrix}
        \hat{1} & v \\
        v^\dagger & -\hat{1}
    \end{pmatrix},
\end{eqnarray}
with $v = i\sigma_z \tau_x$.
For this basis, the BdG Hamiltonian is off diagonal,
\begin{eqnarray}
    V_\Gamma H(\bm{k})V_\Gamma^\dagger = \begin{pmatrix}
        & q(\bm{k}) \\
        q^\dagger(\bm{k}) &
    \end{pmatrix}, 
\end{eqnarray}
with,
\begin{eqnarray}
    q(\bm{k}) = H_nv - \Delta.
\end{eqnarray}
Using $q(\bm{k})$, the 1D crystalline winding number~(\ref{eq:sym winding num}) can be written as
\begin{eqnarray}
    \label{eq:ref1}
    w_{\mathcal{C}_{a} }(\bm{k}_{a\text{-axis}}) &=& -\frac{1}{4\pi i}\int_{-2\pi}^{2\pi} dk_a
    \tr [\Gamma_{\mathcal{C}_{a} } H^{-1}\partial_{k_a}H] \\
    &=& \frac{1}{2\pi} \mathrm{Im} \biggr[ \int_{-2\pi}^{2\pi} dk_a \partial_{k_a} 
    \mathrm{ln}(\mathrm{det~q(\bm{k_a})}) \biggr],
    \label{eq:a7}
\end{eqnarray}
where $\bm{k}_{a\text{-axis}}= (k_b,k_c) = (0,0)$, or  $(\pi,\pi)$. 
We evaluate (\ref{eq:a7}) for the two-orbital model given in Sec. IV,
\begin{eqnarray}
    H(k_a) = \xi(k_a) + f_{x}(k_a)\tau_x + f_y(k_a)\tau_y 
\end{eqnarray}
where $\xi(\bm{k}) = \varepsilon_0(\bm{k})-\mu$ .
We consider both intraorbital and interorbital pairings,
\begin{eqnarray}
    \Delta = \begin{pmatrix}
        \Delta_{11} & \Delta_{12} \\
        \Delta_{21} & \Delta_{22} 
    \end{pmatrix},
\end{eqnarray}
where the indices $1$, $2$ label orbital degrees of freedom.
For simplicity we assume $\Delta_{11}=\Delta_{22}$, because the components satisfying $\Delta_{11}=-\Delta_{22}$ are spin-singlet pairings.
In fact, we have examined that spin-singlet pairings do not affect the 1D crystalline winding number.
Considering spin-triplet pairings, we can write
\begin{eqnarray}
    \Delta_{11} &=& \Delta_{22} = i \bm{d}_{11}^{\mathrm{ot}} \cdot\boldsymbol{\sigma}\sigma_y, \\
    \Delta_{12} &=& i(\bm{d}_{12}^{\mathrm{ot}} + \bm{d}_{12}^{\mathrm{os}})\cdot\boldsymbol{\sigma}\sigma_y, \\
    \Delta_{21} &=& i(\bm{d}_{12}^{\mathrm{ot}} - \bm{d}_{12}^{\mathrm{os}})\cdot\boldsymbol{\sigma}\sigma_y,
\end{eqnarray}
where $d^{\rm ot}$ ($d^{\rm os}$) is the $d$-vector of orbital-triplet (orbital-singlet) pairings.
In the case of the $A_u$ state, on the $\mathcal{C}_a$ rotation axis only $d_a^{\rm ot}$ and $d_c^{\rm os}$
are nonzero.
For this model we obtain
\begin{eqnarray}
    \mathrm{det}~q(k) &=& [\xi^2-|f|^2 +d_{11}^2 - d_{12}^2 + \Delta^{e2} \nn \\
    &&\hspace{50pt} + 2i(\xi d_{12} - f_x d_{11})]^2 \nn \\
    &\equiv& (A + iB)^2,
\end{eqnarray}
with,
\begin{eqnarray}
    A &=& \xi^2-|f|^2 +d_{11}^2 - d_{12}^2 + \Delta^{e2}, \\
    B &=& 2(\xi d_{12} - f_x d_{11}),
\end{eqnarray}
where $f = f_{x} - i f_y$, 
and we renamed $d^{\rm ot}_a = d$ and $d^{\rm os}_c = i\Delta^e$.
Using this we rewrite Eq.~(\ref{eq:a7}) as
\begin{eqnarray}
    w_{\mathcal{C}_{a} }(\bm{k}_{a\text{-axis}}) = -\frac{1}{\pi} \int d_{k_a}\epsilon^{xy}m_x\partial_{k_a}m_y,
    \label{eq:mm}
\end{eqnarray}
with
\begin{eqnarray}
    m_1 = \frac{B}{A^2 + B^2}, ~ m_2 = \frac{A}{A^2 + B^2}.
\end{eqnarray}
Since topological invariants are not affected by continuously changing system parameters as long as singular points of the integrand are avoided,
we can rescale $d$ as $\alpha d ~(\alpha\rightarrow 0)$,
except the neighborhoods of the zero of $A$, arriving at
\begin{eqnarray}
    &&m_1 \rightarrow 0, \nn \\
    &&m_2 \rightarrow \mathrm{sgn}(A). \nn
\end{eqnarray}
This means that the integral (\ref{eq:mm}) is dominated by contributions from the neighborhood of the zero of $A$, which can be evaluated by
expanding $A$ and $B$ around $k_{a0}$ satisfying $A(k_{a0})=0$,
\begin{eqnarray}
    A(k_a) &=& \partial_{k_a}[\xi^2 - |f|^2](k_a-k_{a0}) + \cdots, \\
    B(k_a) &=& 2(\xi d_{12}(k_{a0}) - f_x d_{11}(k_{a0})) + \cdots.
\end{eqnarray}
Then we arrive at
\begin{eqnarray}
    w_{\mathcal{C}_{a} } &=& \sum_{\xi^2 - |f|^2 + \Delta^{e2}= 0}
    \mathrm{sgn}[\partial_{k_a}(\xi^2 - |f|^2)] \nn \\
    &&\hspace{60pt}\times \mathrm{sgn}[\xi d_{12} - f_x d_{11}].
    \label{eq:wca1}
\end{eqnarray}
We can generally assume that $\Delta^e$ is much smaller than the bandwidth and hence negligible in (\ref{eq:wca1}).
The Hamiltonian in the normal state $H_n$ has two energy bands,
$E_{\pm}(\bm{k}_F) = \xi(\bm{k}_F) \pm |f(\bm{k}_F)|$.
We assume that the Fermi level crosses the lower band and is sufficiently far away from the upper band.
Then the sum is taken on the Fermi surfaces, which results in
\begin{eqnarray}
    \mathrm{sgn}[\partial_{k_a}(\xi^2 - |f|^2)] 
    \sim \mathrm{sgn}[\partial_{k_a}E_-].
\end{eqnarray}
Then we end up with
\begin{eqnarray}
    w_{\mathcal{C}_{a}} = \sum_{E(\bm{k}_F)=0}\mathrm{sgn}[\xi d_{12} - f_x d_{11}]\mathrm{sgn}[\partial_{k_a}E_-].
    \label{eq:wna}
\end{eqnarray}
This is the Fermi surface formula of $w_{\mathcal{C}_{a}} $.
In the main text we consider only interorbital pairings, and hence (\ref{eq:wna}) is rewritten as
\begin{eqnarray}
    w_{\mathcal{C}_{a}} = \sum_{E(\bm{k}_F)=0}\mathrm{sgn}[(\bm{d}_{12})_ a]\mathrm{sgn}[\partial_{k_a}E_-].
    \label{eq:fsf_ca}
\end{eqnarray}
In the calculation of $w_{\mathcal{C}_a}$, we use this formula.
The Fermi surface formulas for other 1D crystalline winding numbers can be derived in a similar manner. 
Note that in Eq.~(\ref{eq:wna}), the orbital-singlet gap $\Delta^e$ is not important.
On the other hand, however, the 1D crystalline winding number associated with two fold rotation around the $c$-axis
is given by
\begin{eqnarray}
    w_{\mathcal{C}_{c}} = \sum_{E(\bm{k}_F)=0}\mathrm{sgn}[f_x (\bm{d}_{12})_c - f_y \Delta^e]\mathrm{sgn}[\partial_{k_a}E_-],
    \label{eq:wcc}
\end{eqnarray}
which crucially depends on $\Delta^e$.
The dependence on $\Delta^e$ similar to (\ref{eq:wcc}) also appears in the Fermi surface formulas of the 1D winding number associated with 
$\mathcal{M}_{bc}$ for the $B_{2u}$ state,
and $\mathcal{M}_{ca}$ for the $B_{3u}$ state.
For other cases we can safely ignore $\Delta^e$ in the calculation of the 1D crystalline winding number.
\begin{figure}[t]
    \includegraphics[width=\linewidth]{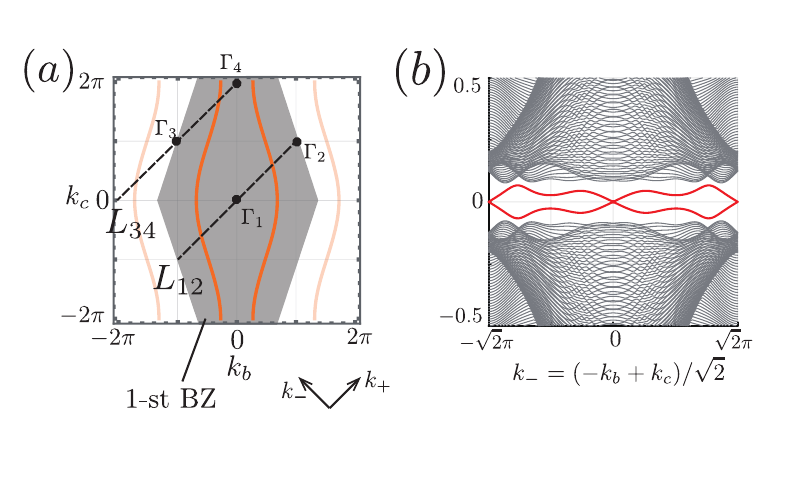}
    \caption{(a) The 2D first BZ for $k_a$ = 0 (shaded area). 
    The dashed lines connect time-reversal-invariant momenta.
    Weak $\mathbb{Z}_2$ invariants can be defined in this 1D subspace.
    (b) Numerical results of quasiparticle energy spectra of the $A_u$ state for an open surface perpendicular to the $(011)$ direction.
    Surface MZMs appear at $k_- = 0$ and $\sqrt{2}\pi$.}
    \label{fig:011}
\end{figure}

\section{1D topological invariant of class D$\mathrm{I}\hspace{-1.2pt}\mathrm{I}\hspace{-1.2pt}\mathrm{I}$}
A single crystal of UTe$_2$ is easily cleaved at the $(011)$ surface.
Thus it is useful to investigate Majorana surface states on an open surface perpendicular to the $(011)$ direction.
In this Appendix, we present some results on this issue.
It is noted that on the $(011)$ surface, the point-group symmetry is not preserved, and hence surface MZMs on this surface, if they exist, are not protected by
crystalline symmetry.  However, weak $\mathbb{Z}_2$ invariants protected by TRS may be nontrivial.
In Fig.~\ref{fig:011}(a) we show the 2D BZ at $k_a = 0$, where there are four time-reversal-invariant momenta ($\Gamma_{1-4}$).
Then we can introduce a weak $\mathbb{Z}_2$ topological invariant
on a time-reversal-invariant loop $L_{ij}$ which connects $\Gamma_i$ and $\Gamma_j$ ($i\neq j$): 
\begin{eqnarray}
    \nu[L_{ij}] = \frac{1}{\pi}\oint_{L_{ij}} dk_+ \mathcal{A}^-(k) ~~~~~ (\mathrm{mod}~2)
\end{eqnarray}
The topological invariants are determined only by the topology of the Fermi surfaces:
$\nu[L_{ij}]$ is nontrivial (trivial) when there are odd (even) numbers of the Fermi surfaces crossing $L_{ij}$
between $\Gamma_i$ and $\Gamma_j$.
Therefore $\nu[L_{12}]$ and $\nu[L_{34}]$ are nontrivial, 
and thus surface MZMs appear on the $(011)$ surface.
In Fig.~\ref{fig:011}(b) we show the numerical results of the energy spectra for the $A_u$ state
with the  $(011)$ open surface.
Surface MZMs appear at $k_- = (-k_b + k_c)/\sqrt{2} = 0,~ \sqrt{2}\pi$.
If the hole Fermi surface is taken into account,
similar surface MZMs appear on the $(101)$ surface.
However, as mentioned above, the topological invariants are not protected by crystalline symmetry
because the 1D subspace we consider is not any symmetry line.
Thus these MZMs are unstable against perturbations compared to those protected by
the 1D crystalline winding number or the mirror Chern number discussed in the main text.
\begin{figure}[t!]
    \includegraphics[width=\linewidth]{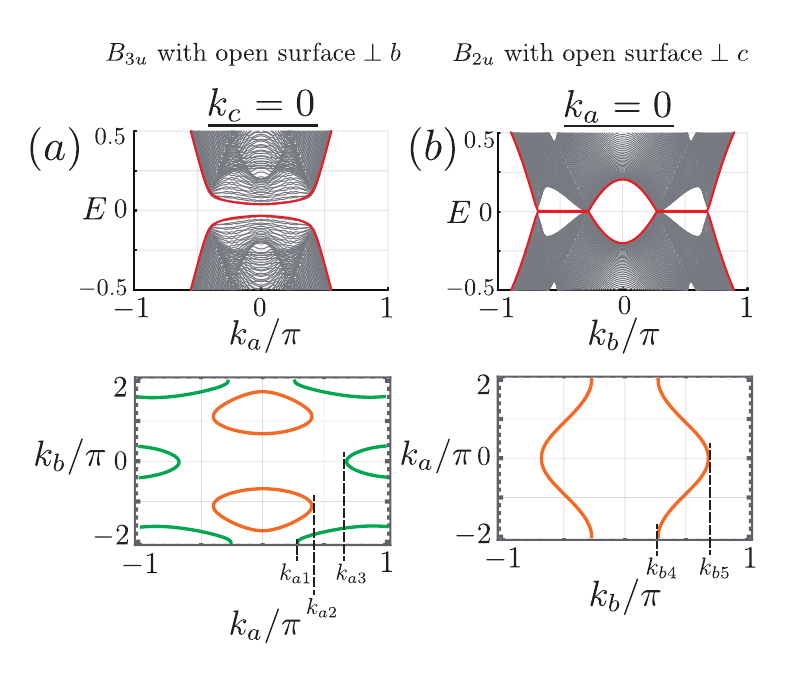}
    \caption{(a) Top: Quasiparticle energy spectra vs $k_a$ for the $B_{3u}$ state with an open surface perpendicular to 
    the $a$-axis. $k_c$ is set to be zero.
    Bottom: The cross section of the electron (orange) and hole (green) Fermi surfaces at $k_c=0$.  
    (b) Top: Quasiparticle energy spectra vs $k_b$ for the $B_{2u}$ state with an open surface perpendicular to 
    the $c$ axis.  $k_a$ is set to be zero. Bottom: The cross section of the electron Fermi surfaces at $k_a=0$. 
}
    \label{fig:ad2}
\end{figure}

\section{Numerical results for other basis functions}
In the main text we have used basis functions of the form $\sin k_{\mu}$ ($\mu=a,b,c$) for the $\bm{d}$ vector of orbital-triplet pairings 
in all calculations.
The crystal symmetry $Immm$ allows other types of basis functions such as $\sin k_a/2 \cos k_b/2 \cos k_c/2$.
Here we show some numerical results of surface Majorana states, taking this point into account. 
We consider the $\bm{d}$ vector of the $B_{3u}$ state of the form
\begin{eqnarray}
    d_{B_{3u}} = \begin{pmatrix}
        0 \\
        C_1 \sin k_c + C_3 \sin k_c/2 \cos k_a/2 \cos k_b/2 \\
        C_2 \sin k_b + C_4 \sin k_b/2 \cos k_a/2 \cos k_c/2
    \end{pmatrix}.
\end{eqnarray}
We set the parameters as $C_1 = C_2 = 0.1, C_3 = C_4 = 0.2$ for numerical calculations.
In this case, the 1D crystalline winding number associated with $\mathcal{M}_{ab}$ symmetry is zero
and hence the flat band of surface MZM found in Sec. V D disappears [see Fig.~\ref{fig:ad2}(a)].
However, if one takes into account the hole Fermi surface, the Majorana zero-energy flat band appears for $k_{a1}<|k_a|<k_{a3}$ 
on the surface perpendicular to the $b$ axis. [The definitions of $k_{a1}$ and $k_{a3}$ are shown in Fig.~\ref{fig:ad2}(a).]
Moreover, we also find surface MZMs on the surface perpendicular to the $c$ axis.

On the other hand, for the  $B_{2u}$ state with the $\bm{d}$ vector,
\begin{eqnarray}
    d_{B_{2u}} = \begin{pmatrix}
        C_1 \sin k_c + C_3 \sin k_c/2 \cos k_a/2 \cos k_b/2 \\
        0 \\
        C_2 \sin k_a + C_4 \sin k_a/2 \cos k_b/2 \cos k_c/2
    \end{pmatrix},
\end{eqnarray}
the Majorana flat band found in Sec. V C partly survives as shown in Fig.~\ref{fig:ad2}(b), and also,
the corresponding 1D crystalline winding number for $k_{b4}<k<k_{b5}$ is nonzero.
In a similar manner we examined that for all IRs of pairing states, surface MZMs appear for other forms of basis functions.

\end{appendix}

\bibliography{paper}

\end{document}